\documentclass[12pt]{article}

\usepackage{latexsym}
\usepackage{amsmath}
\usepackage{verbatim}
\usepackage{amssymb}
\usepackage{mathrsfs}
\usepackage{bbold}
\usepackage{slashed}

\usepackage[
top    = 2.50cm,
bottom = 2.50cm]{geometry}

\catcode`\@=11
\def\marginnote#1{}
\newcount\hour
\newcount\minute
\newtoks\amorpm
\hour=\time\divide\hour by60
\minute=\time{\multiply\hour by60 \global\advance\minute by-\hour}
\edef\standardtime{{\ifnum\hour<12 \global\amorpm={am}%
        \else\global\amorpm={pm}\advance\hour by-12 \fi
        \ifnum\hour=0 \hour=12 \fi
        \number\hour:\ifnum\minute<10 0\fi\number\minute\the\amorpm}}
\edef\militarytime{\number\hour:\ifnum\minute<10 0\fi\number\minute}
\def\draftlabel#1{{\@bsphack\if@filesw {\let\thepage\relax
   \xdef\@gtempa{\write\@auxout{\string
      \newlabel{#1}{{\@currentlabel}{\thepage}}}}}\@gtempa
   \if@nobreak \ifvmode\nobreak\fi\fi\fi\@esphack}
        \gdef\@eqnlabel{#1}}
\def\@eqnlabel{}
\def\@vacuum{}
\def\draftmarginnote#1{\marginpar{\raggedright\scriptsize\tt#1}}
\def\draft{\oddsidemargin -.5truein
        \def\@oddfoot{\sl preliminary draft \hfil
        \rm\thepage\hfil\sl\today\quad\militarytime}
        \let\@evenfoot\@oddfoot \overfullrule 3pt
        \let\label=\draftlabel
        \let\marginnote=\draftmarginnote
   \def\@eqnnum{(\theequation)\rlap{\kern\marginparsep\tt\@eqnlabel}%
\global\let\@eqnlabel\@vacuum}  }

\def\preprint{\twocolumn\sloppy\flushbottom\parindent 1em
        \leftmargini 2em\leftmarginv .5em\leftmarginvi .5em
        \oddsidemargin -.5in    \evensidemargin -.5in
        \columnsep 15mm \footheight 0pt
        \textwidth 250mmin      \topmargin  -.4in
        \headheight 12pt \topskip .4in
        \textheight 175mm
        \footskip 0pt
        \def\@oddhead{\thepage\hfil\addtocounter{page}{1}\thepage}
        \let\@evenhead\@oddhead \def\@oddfoot{} \def\@evenfoot{} }

\def\titlepage{\@restonecolfalse\if@twocolumn\@restonecoltrue\onecolumn
     \else \newpage \fi \thispagestyle{empty}\c@page\z@ 
        \def\thefootnote{\fnsymbol{footnote}} }

\def\endtitlepage{\if@restonecol\twocolumn \else  \fi
        \def\thefootnote{\arabic{footnote}}
        \setcounter{footnote}{0}}  %\c@footnote\z@ }

\catcode`@=12
\relax
\def\bea{\begin{array}}
\def\bem{\begin{displaymath}}
\def\beq{\begin{equation}}

\def\eea{\end{array}}
\def\eem{\end{displaymath}}
\def\eeq{\end{equation}}
\def\s2w{\sin^2 \theta_W}

\relax
\def\crbig{\\\noalign{\vspace {3mm}}}

\newcommand{\skipthispart}[1]{}
\relax
\begin{document}
%\topmargin-2.4cm
%\textwidth =450pt
%\textheight 10in
%\draft
%\preprint
%
%
%
\begin{titlepage}
\begin{flushright}
\today
\end{flushright}
\vspace{0.7cm}

\begin{center}{\Large\bf
The effective supergravity of Little String Theory}

\vspace{1.0cm}

{\large\bf Ignatios Antoniadis $^{1,2}$\,\footnote{antoniad@lpthe.jussieu.fr}, 
Antonio Delgado $^{3}$\,\footnote{antonio.delgado@nd.edu},\\
Chrysoula Markou $^1$\,\footnote{chrysoula@lpthe.jussieu.fr},  
Stefan Pokorski $^{4}$\,\footnote{Stefan.Pokorski@fuw.edu.pl}
}

\vspace{6mm}

$^1$ LPTHE, UMR CNRS 7589, Sorbonne Universit\'es,  \\ UPMC Paris 6, 75005 Paris, France

\vspace{6mm}

$^2$ Albert Einstein Center for Fundamental Physics \\
Institute for Theoretical Physics, University of Bern \\
Sidlerstrasse 5, CH--3012 Bern, Switzerland \\

\vspace{6mm}

$^3$ Department of Physics, University of Notre Dame\\
225 Nieuwland hall, Notre Dame, Indiana 46556, USA\\

\vspace{6mm}

$^4$ Institute of Theoretical Physics, Faculty of Physics\\
University of Warsaw, ul. Pasteura 5, PL-02-093 Warsaw, Poland

\end{center}
\vspace{0.5cm}

\begin{center}
{\large\bf Abstract}
\end{center}
\begin{quote}
In this work we present the minimal supersymmetric extension of the five-dimensional dilaton-gravity theory that captures the main properties of the holographic dual of little string theory. It is described by a particular gauging of $\mathcal{N}=2$ supergravity coupled with one vector multiplet associated to the string dilaton, along the $U(1)$ subgroup of $SU(2)$ R-symmetry. The linear dilaton in the fifth coordinate solution of the equations of motion (with flat string frame metric) breaks half of the supersymmetries to $\mathcal{N}=1$ in four dimensions. The non-supersymmetric version of this model was found recently as a continuum limit of a discretised version of the so-called clockwork setup.
\end{quote}

\end{titlepage}

\newpage
\section{Introduction}

Besides its own theoretical interest, Little String Theory provides a framework with interesting phenomenological consequences. On one hand, it offers a way to address the hierarchy when the string scale is at the TeV scale~\cite{Antoniadis:2011qw, Baryakhtar:2012wj, Cox:2012ee}, without postulating large extra dimensions (in string units) but instead an ultra-weak string coupling~\cite{Antoniadis:1999rm, Antoniadis:2001sw}. On the other hand, LST appeared recently as a continuum limit of the so-called clockwork models~\cite{Giudice:2016yja} which address the hierarchy in an apriori different way~\cite{Choi:2015fiu, Kaplan:2015fuy}.

Little String Theory (LST) corresponds to a non-trivial weak coupling limit of string theory in six dimensions with gravity decoupled and is generated by stacks of (Neveu-Schwarz) NS5-branes~\cite{Aharony:2004xn}. Its holographic dual corresponds to a seven-dimensional gravitational background with flat string-frame metric and the dilaton linear in the extra dimension~\cite{Aharony:1998ub}. Its properties can be studied in a simpler toy model by reducing the theory in five dimensions. Introducing back gravity weakly coupled, one has to compactify the extra dimension on an interval and place the Standard Model on one of the boundaries, in analogy with the Randall-Sundrum model~\cite{Randall:1999ee} on a slice of a five-dimensional (5d) anti-de Sitter bulk~\cite{Antoniadis:2011qw}.

Since we know that the bulk LST geometry preserves space-time supersymmetry, in this work we study the corresponding effective supergravity which in the minimal case is $\mathcal{N}=2$. In principle, there should be a generalisation with more supersymmetries, or equivalently in higher dimensions. The $\mathcal{N}=2$ gravity multiplet  contains the graviton, a graviphoton and the gravitino (8 bosonic and 8 fermionic degrees of freedom), while the heterotic (or type I) string dilaton is in a vector multiplet containing a vector, a real scalar and a fermion. The corresponding supergravity action~\cite{Gunaydin:1984pf} admits a gauging of the $U(1)$ subgroup of the $SU(2)$ R-symmetry, that generates a potential for the single scalar field~\cite{Gunaydin:1984pf, Gunaydin:1984ak}. This potential depends on two parameters allowing a multiple of possibilities with critical or non critical points, or even flat potential with supersymmetry breaking. Here, we observe that the vanishing of one of the parameters generates the runaway dilaton potential of the non-critical string. This potential has no critical point with 5d maximal symmetry but it leads to the linear dilaton solution in the fifth coordinate that preserves 4d Poincar\'e symmetry. We show that this solution breaks one of the two supersymmetries, leading to $\mathcal{N}=1$ in four dimensions.

%{\bf Discuss the spectrum}

The outline of the paper is the following. In Section 2, we review the gauged $\mathcal{N}=2$ supergravity in five dimensions, based on the references~\cite{Gunaydin:1983bi, Gunaydin:1984pf, Gunaydin:1984ak}, and specialize in the case of one vector multiplet using the results of the string effective action of ref.~\cite{Antoniadis:1995vz}.
In Section 3, we present the 5d graviton-dilaton toy model that describes the holographic dual of LST and identify it with a particular choice of the gauging of the $\mathcal{N}=2$ supergravity. We also show that the linear dilaton solution preserves half of the supersymmetries, i.e. $\mathcal{N}=1$ in four dimensions. In Section 4, we write the complete Lagrangian, including the fermion terms, depending on three constant parameters. In Section 5, we derive the spectrum classified using the 4d Poincar\'e symmetry and we conclude with some phenomenological remarks. Finally, there are three appendices containing our conventions, the equations of motion with the linear dilaton solution, and some explicit calculations that we use in the study of supersymmetry transformations.

\section{Gauged $\mathcal{N}=2$, $D=5$ Supergravity} \label{Secg}

The references used in the following are \cite{Gunaydin:1984pf}, \cite{Gunaydin:1983bi} and \cite{Gunaydin:1984ak}, while our conventions may be found in the Appendix (\ref{AppA}). In $D=5$ spacetime dimensions, the pure $\mathcal{N}=2$ supergravity multiplet contains the graviton $e_M^m$, the gravitino $SU(2)$--doublet $\psi_M^i$, where $i$ is the $SU(2)$ index, and the graviphoton, while the $\mathcal{N}=2$ Maxwell multiplet contains a real scalar $\phi$, an $SU(2)$ fermion doublet $\lambda^i$ and a gauge field. Upon coupling $n$ Maxwell multiplets to pure $\mathcal{N}=2$, $D=5$ supergravity, the total field content of the coupled theory can be written as
\begin{equation}
\{ e_M^m \,,\, \psi_M^i \,,\, A_M^I \,,\, \lambda^{ia} \,,\, \phi^x \} \,,
\end{equation}
where $I=0,1,\dots,n\,$, $a=1,\dots,n$ and $x=1,\dots,n$.
The real scalars $\phi^x$ can be seen as coordinates of an $n$--dimensional space $\mathcal{M}$ that has metric $g_{xy}\,$ that is symmetric for our purposes, while the spinor fields $\lambda^{ia}$ transform in the $n$--dimensional representation of $SO(n)$, which is the tangent space group of $\mathcal{M}$, so that
\beq
g_{xy} = f_x^a\, f_y^b \, \delta_{ab}\,,
\eeq
where $f_x^a$ is the corresponding vielbein. The bosonic part of the Lagrangian is 
\beq
\begin{array}{rcl}
e^{-1} \mathcal{L}_{bos} &=& - \frac{1}{2} \mathcal{R}(\omega) - \frac{1}{2} g_{xy} (\partial_M \phi^x) (\partial^M \phi^y) - \frac{1}{4} G_{IJ} F_{MN}^I F^{MN J}
\crbig
&& + \frac{e^{-1}}{6 \sqrt 6} C_{IJK}\epsilon^{MNP \Sigma \Lambda} F_{MN}^I F_{P \Sigma}^J A_\Lambda^K\,,
\end{array}
\eeq
where $e= \det(e_M^m)$, $\omega$ is the spacetime spin--connection, $G_{IJ}$ is the symmetric gauge kinetic metric, $C_{IJK}$ are totally symmetric constants and the gravitational coupling $\kappa$ has been set equal to $1$. The supersymmetry transformations of the fermions of the theory are 
\beq
\begin{array}{rcl}
\delta \psi_{M i} &=& D_{M}( \omega) \epsilon_i + \dots
\crbig
\delta \lambda_i^a &=& - \frac{1}{2} i f_x^a (\slashed{\partial} \phi^x) \epsilon_i + \dots \,, 
\end{array}
\eeq
where $\epsilon_i$ is the supersymmetry spinor parameter and the dots stand for terms that vanish in the vacuum. 

In fact, the $n$--dimensional $\mathcal{M}$ can be seen as a hypersurface of an $(n+1)$--dimensional space $\mathcal{E}$ with coordinates
\beq
\xi^I = \xi^I (\phi^x, \mathcal{F})\,,
\eeq
where $\mathcal{F}$ is the additional coordinate of $\mathcal{E}$ compared to $\mathcal{M}$. It can be shown that $\mathcal{F}$ is a homogeneous polynomial of degree three and, more precisely, that
\beq
\mathcal{F} = \beta^3 C_{IJK} \xi^I \xi^J \xi^K\,,
\eeq
where $\beta = \sqrt{2/3}$. It can also be shown that, on $\mathcal{M}$, the scalars $\phi^x$ satisfy the constraint
\beq \label{constr}
\mathcal{F} =1 \,.
\eeq
Moreover,
\beq
G_{IJ} = - \frac{1}{2} \partial_I \partial_J \ln \mathcal{F} |_{\mathcal{F}=1} \quad , \quad g_{xy} = G_{IJ} \, \partial_x \xi^I \partial_y \xi^J |_{\mathcal{F}=1} \,,
\eeq
where $\partial_I = \frac{\partial}{\partial \xi^I}$ and $\partial_x = \frac{\partial}{\partial \phi^x}$. Finally, we note that the symmetric third--rank tensor $T_{xyz}$ on $\mathcal{M}$ is covariantly constant for the symmetric $\mathcal{M}$ that we will be concerned with and thus satisfies the algebraic constraint
\beq \label{third}
T_{(xy}^{\hphantom{xy} w} T_{zu)w} = \frac{1}{2} g_{(xy}g_{zu)}\,.
\eeq

The gauging of the $U(1)$ subgroup of $SU(2)$ generates a scalar potential $P$, with
\beq \label{potential}
P = - P_0^2 + P_a P^a \,,
\eeq
where
$P_0$ and $P_a$ are functions of the scalars $\phi^x$ that satisfy the following constraints due to supersymmetry
\beq \label{pot}
\begin{array}{ccl}
P_{0,x} = - \sqrt2 \beta P_x
\crbig
P_{0,x;y} + \beta \, T_{xy}^{\hphantom{xy} z} P_{0,z} - \beta^2 g_{xy} P_0 = 0\,,
\end{array}
\eeq
where the symbols ``$,$'' and ``$;$'' denote differentiation and covariant differentiation respectively and $P_x=f_x^aP_a$. The functions $P_0$ and $P_a$ also appear in the fermion transformations that get deformed due to the gauging, namely
\beq \label{susytr}
\begin{array}{rcl}
\tilde\delta \psi_{M i} &=&  D_M( \omega)  \epsilon_i   +  \frac{ig}{2\sqrt{6}} \, P_0 \, \Gamma_M \varepsilon_{ij}\delta^{jk}\epsilon_k  +\dots
\crbig
\tilde\delta \lambda_i^a &=& - \frac{1}{2} i f_x^a (\slashed{\partial} \phi^x) \epsilon_i  +  \frac{g}{\sqrt{2}}\, P^a \varepsilon_{ij}\delta^{jk}\epsilon_k   +  \dots \,, 
\end{array}
\eeq
where $\tilde{\delta}$ denotes the supersymmetry transformation after the gauging (under which the deformed action is invariant), $g$ is the $U(1)$ coupling constant, $\Gamma_\mu$ is the $\Gamma$--matrix in five spacetime dimensions and the dots stand again for terms that vanish in the vacuum.

Now let us consider the case in which there is only one real physical scalar $s$. In the following, we use $t$ to denote the additional coordinate on $\mathcal{E}$, namely $\xi^I=\xi^I(s, t)\,$, $I=0,1$. The effective supergravity related to the $5$--dimensional model for the gravity dual of LST is given by
\beq
\mathcal{F} = t s^2 + as^3\,,
\eeq
where $a$ is a constant parameter. Indeed in the graviton-dilaton system obtained from string compactifications in five dimensions, the first term corresponds to the tree-level contribution (identifying $t$ with the inverse heterotic string coupling) and the second term to the one-loop correction~\cite{Antoniadis:1995vz}.\footnote{Note a change of notation between $s$ and $t$ compared to Ref.~\cite{Antoniadis:1995vz}.}

The solution of the constraint (\ref{constr}) is then
\beq
t= \frac{1-as^3}{s^2}\,.
\eeq
and the components of the gauge kinetic metric are
\beq \label{gaugec}
G_{tt} = \frac{1}{2} s^4 \quad , \quad G_{s t} = \frac{1}{2} a s^4 \quad , \quad G_{s s} = \frac{1}{s^2} + \frac{1}{2} a^2 s^4 \,.
\eeq
We then find that the scalar metric, the Christoffel symbols and the third--rank tensor (that have only one component each) are respectively
\beq \label{data}
g_{s s} = \frac{3}{s^2}  \quad , \quad  f_s^a= \frac{\sqrt{3}}{s} \quad , \quad \Gamma_{s s}^s = - \frac{1}{ s} \quad , \quad T_{s s s} = \frac{3}{\beta} \frac{1}{s^3}\,,
\eeq
where we have used (\ref{third}) to compute $T_{s s s}$. The system (\ref{pot}) takes thus the form
\beq \label{pot2}
\begin{array}{ccl}
P_{s} = - \frac{\sqrt{3}}{2} P_0'
\crbig
P_0'' + \frac{2}{s} P_0' -\frac{2}{s^2} P_0 = 0\,,
\end{array}
\eeq
whose solution is 
\beq \label{pot3}
\begin{array}{ccl}
P_0 = As+B\frac{1}{s^2}
\crbig
P_s = - \frac{\sqrt{3}}{2}(A-2B\frac{1}{s^3}) \quad,\quad P^a = f_s^a g^{s s}P_s = -\frac{A}{2}s+B\frac{1}{s^2}\,.
\end{array}
\eeq
where $A$, $B$ are constant parameters. Using (\ref{potential}) we then find the potential to be
\beq
P = -3A \Big( \frac{A}{4}s^2+B\frac{1}{s}  \Big)
\eeq
so that the kinetic term and the potential for $s$ take the form
\beq
e^{-1} \mathcal{L}_{dilaton} = - \frac{1}{2} \frac{3}{s^2}(\partial_M s)(\partial^M s) +3A \Bigg( \frac{A}{4}s^2+B\frac{1}{s} \Bigg)\,.
\eeq
Upon redefining 
\beq
\sqrt{3} \ln s = \Phi\,,
\eeq
we obtain the Lagrangian for the canonically normalized $\Phi$
\beq \label{eff}
e^{-1} \mathcal{L}_{dilaton} = -\frac{1}{2} (\partial_M \Phi) (\partial^M \Phi) +3g^2A \Bigg( \frac{A}{4}e^{\frac{2}{\sqrt3}\Phi}+Be^{-\frac{1}{\sqrt3}\Phi} \Bigg)\,.
\eeq

\section{The 5D dual of LST}

The holographic dual of six--dimensional Little String Theory can be approximated by a five--dimensional model, in which the Lagrangian in the bulk takes the following form \cite{Antoniadis:2011qw}\footnote{We neglect the remaining spectator five dimensions of the string background which play no role in the properties of the model relevant for our analysis.}
\beq
e^{-1}\mathcal{L}_{LST} = - \widetilde{M}_5^3 \mathcal{R}-\frac{1}{3} (\partial_M \tilde{\Phi}) (\partial^M \tilde{\Phi}) - e^{\frac{2}{3} \frac{\tilde{\Phi}}{\widetilde{M}_5^{3/2}}} \Lambda
\eeq 
in the Einstein frame, where $\tilde{\Phi}$ is the dilaton and $\Lambda$ is a constant. Upon redefining
\beq\label{correspondence}
 \widetilde{\Phi}  = \sqrt{ \frac{3}{2}} \Phi \quad , \quad  \widetilde{M}_5^3 =  \frac{1}{2} M_5^3
\eeq
and setting the gravitational coupling $\kappa$ in five dimensions equal to one ($\kappa^2=1/{M^3_5}$, where $M_5$ is the Planck mass in five dimensions), we obtain the Lagrangian for the canonically normalized dilaton $\Phi$
\beq \label{lstpot}
e^{-1}\mathcal{L}_{LST} = - \frac{1}{2} \mathcal{R}-\frac{1}{2} (\partial_M \Phi) (\partial^M \Phi) - e^{\frac{2}{\sqrt{3}} \Phi} \Lambda\,.
\eeq
We thus observe that the potential that arises from LST is equal to the potential in (\ref{eff}) for a scalar that belongs to a gauged $\mathcal{N}=2, D=5$ Maxwell multiplet coupled to supergravity, upon making the identification
\beq \label{condit}
\frac{3}{4}g^2 A^2 = - \Lambda \quad , \quad B=0\,.
\eeq
We then have
\beq \label{comp}
P_0 = A e^{\frac{1}{\sqrt{3}} \Phi} \quad , \quad P^a = - \frac{A}{2} e^{\frac{1}{\sqrt{3}} \Phi} \,.
\eeq

Moreover, it is known that the dilaton potential in (\ref{lstpot}) exhibits a runaway behaviour and does not have a $5$--dimensional maximally symmetric vacuum, but has a $4$--dimensional Poincar\'e vacuum in the linear dilaton background
\beq\label{dilaton}
\Phi = C y \,,
\eeq
where $y>0$ is the fifth dimension and $C$ a constant parameter. The background bulk metric is then
\beq\label{metric}
ds^2 =e^{-\frac{2}{\sqrt{3}}Cy} (\eta_{\mu \nu}dx^\mu dx^\nu+dy^2)\,,
\eeq
where $\eta_{\mu \nu}$ is the Minkowski metric of four--dimensional space, under the fine--tuning condition (see Appendix \ref{AppB})
\beq \label{fine}
C=\frac{gA}{\sqrt{2}}\,.
\eeq 

To have at least one unbroken supersymmetry, the fermion transformations must vanish in the vacuum for at least one linear combination of the supersymmetry parameters. Using equations (\ref{comp}), the fermion transformations (\ref{susytr}) take the following form on the four--dimensional brane (in the vacuum)\footnote{The details of this calculation are given in the Appendix (\ref{AppC}).}
\beq \label{syst}
\begin{array}{ccl}
\tilde\delta \psi_{\mu i}=\frac{i}{2\sqrt 3} \Gamma_\mu \Big( i C \Gamma^5  \epsilon_{i}  + \frac{gA}{\sqrt2} \varepsilon_{ij}\delta^{jk}\epsilon_k  \Big) 
\crbig
\tilde\delta \lambda_i= -\frac{1}{2} e^{\frac{1}{\sqrt{3}} Cy}  \, \Big(i C \Gamma^5  \epsilon_{i}  + \frac{gA}{\sqrt2}\varepsilon_{ij}\delta^{jk}\epsilon_k  \Big) \,.
\end{array}
\eeq
Upon diagonalizing  the second of equations (\ref{syst}) and using (\ref{fine}), we find that $\mathcal{N}=2$ supersymmetry is partially broken to $\mathcal{N}=1$, with
\beq \label{susyaa}
\tilde{\delta} (\lambda_1 + i\Gamma^5\lambda_2 ) =0 \quad , \quad \tilde{\delta} ( i\Gamma^5\lambda_1 +\lambda_2)\sim i \Gamma^5\epsilon_1+\epsilon_2 \,.
\eeq
We thus identify $\lambda_1 +  i\Gamma^5\lambda_2$ with the fermion residing in a multiplet of the unbroken $\mathcal{N}=1$ supersymmetry and $ i\Gamma^5\lambda_1 + \lambda_2 $ with the Goldstino of the broken $\mathcal{N}=1$ supersymmetry.
To determine the dependence of $\epsilon_{i}$ on $y$, we consider the $5$--th component of the first of the equations (\ref{susytr}) in the vacuum\footnote{The details of this calculation are given in the Appendix (\ref{AppC}).}
\beq\label{susycomb}
\tilde\delta \psi_{5 i}= \partial_5 \epsilon_{i} + \frac{igA}{2 \sqrt{6}}\Gamma_5  \varepsilon_{ij}\delta^{jk}\epsilon_k\,,
\eeq
which gives
\beq
\epsilon_{1}= e^{\frac{C}{2\sqrt{3}}y}  \tilde{\epsilon} \quad , \quad \epsilon_{2}=- e^{\frac{C}{2\sqrt{3}}y}  \, i\Gamma_5\,  \tilde{\epsilon} ,
\eeq
where $ \tilde{\epsilon}$ is a constant symplectic spinor. The above relations are consistent with the direction of the unbroken supersymmetry $\epsilon_2=-i\Gamma_5\epsilon_1$ from eq.~(\ref{susyaa}).

\section{Final Lagrangian}

The Lagrangian of ungauged $\mathcal{N}=2$, $D=5$ supergravity is
\beq
\begin{array}{rcl}
e^{-1} \mathcal{L}&=& - \frac{1}{2} \mathcal{R}(\omega) - \frac{1}{2} g_{xy} (\partial_M \phi^x) (\partial^M \phi^y) - \frac{1}{4} G_{IJ} F_{MN}^I F^{MN J}
\crbig
& & - \frac{1}{2} \bar{\psi}_M^i \Gamma^{MNP}D_N \psi_{Pi} - \frac{1}{2} \bar{\lambda}^{ia}(\slashed D \delta^{ab}+\Omega_x^{ab}\slashed\partial \phi^x) \lambda_i^b
\crbig
& & - \frac{1}{2}i \bar{\lambda}^{ia} \Gamma^M \Gamma^N \psi_{Mi} f_x^a \partial_N \phi^x + \frac{1}{4} h_I^a \bar{\lambda}^{ia} \Gamma^M \Gamma^{\Lambda P} \psi_{M i} F_{\Lambda P}^I
\crbig
& & + \frac{1}{4}i \Phi_{Iab} \bar{\lambda}^{ia} \Gamma^{MN} \lambda_i^b F_{MN}^I  + \frac{e^{-1}}{6 \sqrt 6} C_{IJK}\epsilon^{MNP\Sigma \Lambda} F_{MN}^I F_{P \Sigma}^J A_\Lambda^K 
\crbig
& &  - \frac{3i}{8\sqrt{6}} h_I [\bar{\psi}_M^i \Gamma^{MNP\Sigma} \psi_{Ni}F_{P\Sigma}^I+2 \bar{\psi}^{Mi}\psi_i^N F_{MN}^I]
\crbig
& &+ \textrm{\,(4--fermion terms)}\,,
\end{array}
\eeq
where $\Omega_x^{ab}$ is the spin--connection of the scalar manifold and $h_I$, $h_I^x$ and $\Phi_{Ixy}$ are functions of the scalars that will be defined later.

Upon gauging $U(1)$, the Lagrangian aquires the additional terms
\beq
\begin{array}{rcl}
e^{-1} \mathcal{L}'&=&  -g^2P - \frac{i\sqrt{6}}{8} g \bar{\psi}^i_M \Gamma^{MN} \psi_N^j \delta_{ij} P_0
\crbig
&& - \frac{g}{\sqrt{2}} \bar{\lambda}^{ia} \Gamma^M \psi_M^j \delta_{ij} P_a + \frac{ig}{2\sqrt{6}} \bar{\lambda}^{ia} \lambda^{jb} \delta_{ij} P_{ab}\,,
\end{array}
\eeq
and the derivatives become
\beq
D_M \lambda^{ia}+ \Omega_x^{ab}\partial_M \phi^x \lambda^{bi} \Rightarrow (\tilde{\mathcal{D}}_M \lambda^a)^i \equiv D_M \lambda^{ia}  + \Omega_x^{ab}\partial_M \phi^x \lambda^{bi} + g\upsilon_I A_M^I \delta^{ij} \lambda_j^a\,,
\eeq
where $\upsilon_I$ is an arbitrary constant vector and
\beq
P_{ab} \equiv \frac{1}{2}\delta_{ab}P_0 + 2\sqrt{2} T_{abc} P^c\,.
\eeq
Using (\ref{data}) and (\ref{comp}) we find that for a single scalar
\beq
P_{aa} = \frac{1}{2} P_0 + 2\sqrt{2} (f_s^a)^{-3} T_{sss} P^a =-  \frac{A}{2} e^{\frac{1}{\sqrt{3}}\Phi}\,.
\eeq
Consequently,
\beq
\begin{array}{rcl}
e^{-1} \mathcal{L}'&=&  \frac{3g^2A^2}{4}e^{\frac{2}{\sqrt3}\Phi}  - \frac{i\sqrt{6}}{8} g A \, e^{\frac{1}{\sqrt3}\Phi}   \, \bar{\psi}^i_M \Gamma^{MN} \psi_N^j \delta_{ij}
\crbig
&& + \frac{gA}{2\sqrt{2}} \,  e^{\frac{1}{\sqrt3}\Phi}   \, \bar{\lambda}^{i} \Gamma^M \psi_M^j \delta_{ij} - \frac{igA}{4\sqrt{6}} \, e^{\frac{1}{\sqrt3}\Phi}   \, \bar{\lambda}^{i} \lambda^{j} \delta_{ij} \,.
\end{array}
\eeq

In addition, after the gauging, the following equations hold \cite{Gunaydin:1984pf}
\beq
P_0 = 2h^I \upsilon_I \quad  , \quad P^a = \sqrt{2} h^{Ia} \upsilon_I \,,
\eeq
so using (\ref{comp}) we find that
\beq
h^I= \frac{A}{2} \upsilon^I e^{\frac{1}{\sqrt{3}}\Phi} \quad , \quad h^{Ia}= - \frac{A}{2 \sqrt{2}}\upsilon^I e^{\frac{1}{\sqrt{3}}\Phi} \,,
\eeq
where we have assumed that $\upsilon^I \upsilon_I=1$ for simplicity. It thus follows that
\beq
h_I \equiv G_{IJ} h^J =  \frac{A}{2}  G_{IJ} \upsilon^J e^{\frac{1}{\sqrt{3}}\Phi}  \quad , \quad h_I^a \equiv G_{IJ} h^{Ia} = - \frac{A}{2 \sqrt{2}}G_{IJ}\upsilon^J e^{\frac{1}{\sqrt{3}}\Phi}  \,,
\eeq
where we have used the fact that $G_{IJ}$ raises and lowers $I,J$ indices.
Moreover,
\beq
\Phi_{Iab}\equiv \Phi_{Ixy} f_a^x f_b^y \equiv \sqrt{\frac{2}{3}} \Big(\frac{1}{4} g_{xy}h_I + T_{xyz} h_I^z\Big) f_a^x f_b^y\,,
\eeq
using which we find that for a single scalar
\beq
\Phi_{Iaa}=- \frac{A}{8} \sqrt{\frac{2}{3}} G_{IJ} \upsilon^J e^{\frac{1}{\sqrt{3}}\Phi}\,.
\eeq

Using (\ref{gaugec}), we find that the final Lagrangian $\tilde{\mathcal{L}}  = \mathcal{L} + \mathcal{L}'$ takes the form
\beq
\begin{array}{rcl}
e^{-1} \tilde{\mathcal{L}} &=& - \frac{1}{2} \mathcal{R}(\omega) -\frac{1}{2} (\partial_M \Phi) (\partial^M \Phi)  - \frac{1}{8} e^{\frac{4}{\sqrt{3}}\Phi} F_{MN}^0 F^{MN 0}- \frac{1}{4}a e^{\frac{4}{\sqrt{3}}\Phi} F_{MN}^0 F^{MN 1}
\crbig
& &- \frac{1}{4} ( e^{-\frac{2}{\sqrt{3}}\Phi} + \frac{1}{2}a^2 e^{\frac{4}{\sqrt{3}}\Phi} )F_{MN}^1 F^{MN 1}  - \frac{1}{2} \bar{\psi}_M^i \Gamma^{MNP}\mathcal{D}_N \psi_{Pi} - \frac{1}{2} \bar{\lambda}^{i}\tilde{\slashed {\mathcal{D}}}  \lambda_i
\crbig
& & - \frac{i}{2} (\partial_N\Phi) \, \bar{\lambda}^{i} \Gamma^M \Gamma^N \psi_{Mi} - \frac{A\tilde{\upsilon}}{16\sqrt{2}}\, e^{\frac{5}{\sqrt{3}}\Phi}\, \bar{\lambda}^{i} \Gamma^M \Gamma^{\Lambda P} \psi_{M i} \, F_{\Lambda P}^0
\crbig
& &  - \frac{A}{8\sqrt{2}} \Big(\frac{1}{2}a\tilde{\upsilon}\, e^{\frac{5}{\sqrt{3}}\Phi} + \upsilon^1 e^{-\frac{1}{\sqrt{3}}\Phi} \Big)\, \bar{\lambda}^{i} \Gamma^M \Gamma^{\Lambda P} \psi_{M i} \, F_{\Lambda P}^1 - \frac{iA\tilde{\upsilon}}{64}\sqrt{\frac{2}{3}}e^{\frac{5}{\sqrt{3}}\Phi}\,  \bar{\lambda}^{i} \Gamma^{MN} \lambda_i \, F_{MN}^0 
\crbig
& &- \frac{iA}{32}\sqrt{\frac{2}{3}} \Big(\frac{1}{2}a\tilde{\upsilon}\, e^{\frac{5}{\sqrt{3}}\Phi} + \upsilon^1 e^{-\frac{1}{\sqrt{3}}\Phi} \Big)\,   \bar{\lambda}^{i} \Gamma^{MN} \lambda_i \,  F_{MN}^1  + \frac{e^{-1}}{6 \sqrt 6} C_{IJK}\epsilon^{MNP\Sigma \Lambda} F_{MN}^I F_{P \Sigma}^J A_\Lambda^K 
\crbig
& &  - \frac{3iA\tilde{\upsilon}}{32\sqrt{6}}e^{\frac{5}{\sqrt{3}}\Phi} \,  [\bar{\psi}_M^i \Gamma^{MNP\Sigma} \psi_{Ni}F_{P\Sigma}^0+2 \bar{\psi}^{Mi}\psi_i^N F_{MN}^0]
\crbig
& &  - \frac{3iA}{16\sqrt{6}}  \Big(\frac{1}{2}a\tilde{\upsilon}\, e^{\frac{5}{\sqrt{3}}\Phi} + \upsilon^1 e^{-\frac{1}{\sqrt{3}}\Phi} \Big)\,  [\bar{\psi}_M^i \Gamma^{MNP\Sigma} \psi_{Ni}F_{P\Sigma}^1+2 \bar{\psi}^{Mi}\psi_i^N F_{MN}^1]
\crbig
& &  + \frac{3g^2A^2}{4}e^{\frac{2}{\sqrt3}\Phi}  - \frac{i\sqrt{6}}{8} g A \, e^{\frac{1}{\sqrt3}\Phi}   \, \bar{\psi}^i_M \Gamma^{MN} \psi_N^j \delta_{ij}
\crbig
&& + \frac{gA}{2\sqrt{2}} \,  e^{\frac{1}{\sqrt3}\Phi}   \, \bar{\lambda}^{i} \Gamma^M \psi_M^j \delta_{ij} - \frac{igA}{4\sqrt{6}} \, e^{\frac{1}{\sqrt3}\Phi}   \, \bar{\lambda}^{i} \lambda^{j} \delta_{ij} 
\crbig
&& 
+ \textrm{\,(4--fermion terms)}\,.
\end{array}
\eeq
where $A_M^0$ and $A_M^1$ correspond to the graviphoton and the gauge field of the vector multiplet respectively and we have set $\tilde{\upsilon}=\upsilon^0 + a\upsilon^1$. Since the parameter $A$ appears only through the combination $gA$ in the additional terms $\mathcal{L}'$ induced by the gauging, we choose to set $A=1$. Moreover, at tree--level we may set $a=0$, as discussed in section \ref{Secg}. The final Lagrangian then takes the form
\beq\label{lagrangian}
\begin{array}{rcl}
e^{-1} \tilde{\mathcal{L}} &=& - \frac{1}{2} \mathcal{R}(\omega) -\frac{1}{2} (\partial_M \Phi) (\partial^M \Phi)  - \frac{1}{8} e^{\frac{4}{\sqrt{3}}\Phi} F_{MN}^0 F^{MN 0} - \frac{1}{4}  e^{-\frac{2}{\sqrt{3}}\Phi} F_{MN}^1 F^{MN 1}
\crbig
& &  - \frac{1}{2} \bar{\psi}_M^i \Gamma^{MNP}\mathcal{D}_N \psi_{Pi} - \frac{1}{2} \bar{\lambda}^{i}\tilde{\slashed {\mathcal{D}}}  \lambda_i- \frac{i}{2} (\partial_N\Phi) \, \bar{\lambda}^{i} \Gamma^M \Gamma^N \psi_{Mi}
\crbig
& &  - \frac{\upsilon^0}{16\sqrt{2}}\, e^{\frac{5}{\sqrt{3}}\Phi}\, \bar{\lambda}^{i} \Gamma^M \Gamma^{\Lambda P} \psi_{M i} \, F_{\Lambda P}^0 - \frac{\upsilon^1}{8\sqrt{2}}   e^{-\frac{1}{\sqrt{3}}\Phi} \bar{\lambda}^{i} \Gamma^M \Gamma^{\Lambda P} \psi_{M i} \, F_{\Lambda P}^1 
\crbig
& &  - \frac{i\upsilon^0}{64}\sqrt{\frac{2}{3}}e^{\frac{5}{\sqrt{3}}\Phi}\,  \bar{\lambda}^{i} \Gamma^{MN} \lambda_i \, F_{MN}^0 - \frac{i\upsilon^1}{32}\sqrt{\frac{2}{3}}  e^{-\frac{1}{\sqrt{3}}\Phi}   \bar{\lambda}^{i} \Gamma^{MN} \lambda_i \,  F_{MN}^1 
\crbig
& & + \frac{1}{6 \sqrt 6}\, e^{\frac{5}{\sqrt{3}}\Phi}\,  C_{IJK}\epsilon^{MNP\Sigma \Lambda} F_{MN}^I F_{P \Sigma}^J A_\Lambda^K 
\crbig
& &  - \frac{3i\upsilon^0}{32\sqrt{6}}e^{\frac{5}{\sqrt{3}}\Phi} \,  [\bar{\psi}_M^i \Gamma^{MNP\Sigma} \psi_{Ni}F_{P\Sigma}^0+2 \bar{\psi}^{Mi}\psi_i^N F_{MN}^0]
\crbig
& &  - \frac{3i\upsilon^1 }{16\sqrt{6}}   e^{-\frac{1}{\sqrt{3}}\Phi}  [\bar{\psi}_M^i \Gamma^{MNP\Sigma} \psi_{Ni}F_{P\Sigma}^1+2 \bar{\psi}^{Mi}\psi_i^N F_{MN}^1]
\crbig
& &  + \frac{3g^2}{4}e^{\frac{2}{\sqrt3}\Phi}  - \frac{ig\sqrt{6}}{8}   \, e^{\frac{1}{\sqrt3}\Phi}   \, \bar{\psi}^i_M \Gamma^{MN} \psi_N^j \delta_{ij}
\crbig
&& + \frac{g}{2\sqrt{2}} \,  e^{\frac{1}{\sqrt3}\Phi}   \, \bar{\lambda}^{i} \Gamma^M \psi_M^j \delta_{ij} - \frac{ig}{4\sqrt{6}} \, e^{\frac{1}{\sqrt3}\Phi}   \, \bar{\lambda}^{i} \lambda^{j} \delta_{ij} 
\crbig
&& 
+ \textrm{\,(4--fermion terms)}\,.
\end{array}
\eeq
This Lagrangian has three free parameters: $g$, $\upsilon^0$ and $\upsilon^1$.
\vskip 0.5cm

\section{Spectrum and concluding remarks}

The spectrum of the above model can be decomposed using the 4d Poincar\'e invariance of the linear dilaton vacuum solution and should form obviously $\mathcal{N}=1$ supermultiplets. It is known that every 5d field should give rise to a 4d zero mode and a continuum starting from a mass gap fixed by the linear dilaton coefficient $C=g/\sqrt{2}$. Using the results of Ref.~\cite{Antoniadis:2011qw} and the correspondence (\ref{correspondence}), one finds that the parameter $\alpha$ of~\cite{Antoniadis:2011qw} is given by $\alpha=\sqrt{3}C$ and that the mass gap $M_{\rm gap}$ is 
\beq 
M_{\rm gap}=\frac{\sqrt{3}}{2\sqrt{2}}g\, .
\eeq
The continuum becomes an ordinary discrete Kaluza-Klein (KK) spectrum on top of the mass gap, when the fifth coordinate $y$ is compactified on an interval~\cite{Antoniadis:2011qw}, allowing to introduce the Standard Model (SM) on one of the boundaries. This spectrum is valid for the graviton, dilaton and their superpartners by supersymmetry. Notice that the 5d graviton zero-mode has five polarisations that correspond to the 4d graviton, a KK vector and the radion. For the rest of the fields, special attention is needed because of the gauging that breaks half of the supersymmetry around the linear dilaton solution.

Indeed, one of the 4d gravitini acquires a mass fixed by $g$, giving rise to a massive spin-3/2 multiplet together with two spin-1 vectors. These are the 5d graviphoton and the additional 5d vector that have non-canonical, dilaton dependent, kinetic terms, as one can see from the Lagrangian~(\ref{lagrangian}). Using the background (\ref{dilaton}), (\ref{metric}), one finds that the $y$-dependence of the vector kinetic terms at the end of the first line of~(\ref{lagrangian}) is $\exp{\{\pm\sqrt{3}C\}}$ with the plus (minus) sign corresponding to the 5d graviphoton $I=0$ (extra vector $I=1$). It follows that they both acquire a mass given by the mass gap.

We conclude with some comments on some possible phenomenological implications of the above lagrangian. One has to dimensionally reduce it from $D=5$ to $D=4$, upon compactification of the $y$-coordinate. 
Moreover, one has to introduce the SM, possibly on one of the boundaries, a radion stabilization mechanism and the breaking of the leftover supersymmetry. An interesting possibility is to combine all of them along the lines of the stabilisation proposal of~\cite{Cox:2012ee} based on boundary conditions.

%where the usual decomposition into representations of the 4d Lorentz group will occur. 
There are several possibilities for Dark Matter (DM) candidates in this gravitational sector. There are two gravitini that, upon supersymmetry breaking can recombine to form a Dirac gravitino~\cite{Benakli:2014daa}  
 or remain two different Majorana ones. Depending on the nature of their mass, the exact freeze-out mechanism will be different. There are three possible dark photons $A^0_\mu$, $A^1_\mu$ and the KK $U(1)$ coming from the 5d metric that could also be DM or their associated gaugini could also play a similar role, again depending on the compactification of the extra coordinate, on how supersymmetry breaking is implemented, as well as on the radion stabilisation mechanism. In general there could be a very rich phenomenology in the gravitational sector.
 
 Regarding LHC or FCC phenomenology it is going to depend on how the SM fields are included in this setup, we will leave that to a forthcoming publication~\cite{admp}. In general this theory will have KK massive resonances that could be strongly coupled to the SM in a similar fashion as in Randall-Sundrum~\cite{Randall:1999ee} models.

\section*{Note Added}
After the completion of this work, we received the paper~\cite{Kehagias:2017grx} which contains very similar results.

\section*{Acknowledgements}
The work was partially funded by the Swiss National Science Foundation and by the National Science Foundation under Grant No.~PHY-1520966. I.A. would like the thank Amit Giveon for enlightening discussions. The work of S.P. is partially supported by the National Science Centre, Poland, under research grants DEC-2014/15/B/ST2/02157 and DEC-2015/18/M/ST2/00054.

\renewcommand{\thesection}{\Alph{section}}
\setcounter{section}{0}
\renewcommand{\theequation}{\thesection.\arabic{equation}}

\section{Conventions} \label{AppA}
\setcounter{equation}{0}

Our convention for the five--dimensional Minkowski metric is
\beq
\eta_{mn} = \textrm{diag}(-,+,+,+,+)\,,
\eeq
where $m,n,\dots$ are inert indices and $m=1,\dots,5$. For $\Gamma$--matrices we write
\beq
\Gamma_{mn} \equiv \Gamma_{[m} \Gamma_{n]} \equiv \frac{1}{2} (\Gamma_m \Gamma_n - \Gamma_n \Gamma_m)\,.
\eeq
We also have that
\beq
\Gamma^5=\Gamma_5=i\gamma^5=i\gamma_5\,,
\eeq
where $\gamma^5$ is the standard $\gamma^5$ in four--dimensions, such that in the Dirac representation
\beq
\Gamma^5 = i\gamma^5  =\begin{pmatrix}
  0_{2\times2} & i1_{2\times2}\\
  i1_{2\times2}& 0_{2\times2}  \end{pmatrix}\,.
\eeq
The five--dimensional bulk metric of the LST dual is given by
\beq
g_{MN}= \begin{pmatrix}
  e^{-\frac{2}{\sqrt{3}}Cy}\eta_{\mu \nu} & 0_{4\times1}\\
  0_{1\times4} & e^{-\frac{2}{\sqrt{3}}Cy}\  \end{pmatrix} = e^{-\frac{2}{\sqrt{3}}Cy} \eta_{MN} \,.
\eeq

\section{Einstein equation in $5$D} \label{AppB}
\setcounter{equation}{0}

In our conventions, the Einstein equation takes the form
\beq
G_{MN} = 2 T_{MN}\,,
\eeq
where $G_{MN}$ and $T_{MN}$ are the Einstein and the energy--momentum tensor respectively. Moreover, we have that
\beq
G_{MN} = \frac{3}{2}\Big[\frac{1}{2}\partial_M A \partial_N A+\partial_M \partial_N A - \eta_{MN}\Big(\partial_l \partial^l A - \frac{1}{2} \partial_l A \partial^l A\Big)\Big]\,,
\eeq
where $A=A(y)= \frac{2}{\sqrt{3}}Cy$ in our case. This gives
\beq
G_{55}=\frac{3}{2} \Big( \frac{dA}{dy} \Big)^2= 2C^2\,.
\eeq
In addition,
\beq
T_{MN} = (\partial_M \Phi) (\partial_N \Phi) - g_{MN} \Big(\frac{1}{2} (\partial_K \Phi) (\partial^K \Phi) + e^{\frac{2}{\sqrt{3}} \Phi} \Lambda  \Big)\,,
\eeq
so $T_{55}=\frac{1}{2}C^2-\Lambda$. The Einstein equation $G_{55}=2T_{55}$ then gives 
\beq \label{relat}
C=\frac{gA}{\sqrt{2}}\,,
\eeq
where we have used (\ref{condit}).

\section{Spacetime calculations} \label{AppC}
\setcounter{equation}{0}

In the following $M,N,\dots$ are coordinate indices and $n,m,\dots$ are (inert) frame indices of the five--dimensional spacetime. We have that
\beq
g_{MN} = e_M^m \eta_{mn} e_N^n\,.
\eeq
 The only non--vanishing components of the vielbein $e^m$ are thus
\beq
e_\mu^a=e^{-\frac{1}{\sqrt{3}}Cy} \delta_\mu^a \quad , \quad  e_5^5=e^{-\frac{1}{\sqrt{3}}Cy}\,,
\eeq
where $\mu,\nu,\dots$ are the coordinate and $a,b,\dots$ the frame indices on the four--dimensional brane respectively. Moreover,
\beq
e^{a5}= g^{55} e_5^a =0 \quad , \quad e^{55}=g^{55}e_5^5=e^{\frac{2}{\sqrt{3}}Cy}e_5^5 
\eeq
and
\beq
e^{a\nu} = g^{\nu \kappa} e^a_{\kappa} = e^{\frac{2}{\sqrt{3}}Cy}\eta^{\nu \kappa}e^a_{\kappa}  \quad , \quad e_{\mu b} = \eta_{ab} e_\mu^a\,.
\eeq
Consequently,
\beq
\slashed{\partial} \Phi = (\partial_M \Phi )\Gamma^M =  (\partial_M \Phi)e_m^M \Gamma^m = (\partial_M \Phi)(e_M^m)^{-1} \Gamma^m = C (e_5^5)^{-1}\Gamma^5 =Ce^{\frac{1}{\sqrt{3}}Cy}\Gamma^5 \,.
\eeq
Using the second of the equations (\ref{comp}), the second of the equations (\ref{susytr}) then takes the form (in the vacuum)
\beq
 \tilde\delta \lambda_i= -\frac{1}{2}e^{\frac{1}{\sqrt{3}} Cy}  \,\Big(i C \Gamma^5  \epsilon_{i}  + \frac{gA}{\sqrt2}\varepsilon_{ij}\delta^{jk}\epsilon_k  \Big) \,. 
\eeq

The components of the spacetime spin--connection are given by
\beq
\omega_M^{\hphantom{M}mn}(e)=2e^{[mN}e_{[N,M]}^{\hphantom{N,} n]} + e^{m\Lambda} e^{nP} e_{[\Lambda,P]}^{\hphantom{\Lambda,}l} e_{M l}\,.
\eeq
Consequently,
\beq
\omega_\mu^{\hphantom{\mu}ab}(e) = \Big(- e^{[a5}e_{\mu,5}^{\hphantom{\mu,}b]} + \frac{1}{2}e^{a\Lambda} e^{b5} e_{\Lambda,5}^{\hphantom{\Lambda,}l} e_{\mu l} - \frac{1}{2}e^{b\Lambda} e^{a5} e_{\Lambda,5}^{\hphantom{\Lambda,}l} e_{\mu l}    \Big) =0\,,
\eeq
since $e^{a5}=0 $. Moreover, 
\beq
\begin{array}{rcl}
\omega_\mu^{\hphantom{\mu}a5}(e) &=&  \Big(- e^{[a5}e_{\mu,5}^{\hphantom{\mu,}5]} + \frac{1}{2}e^{a\Lambda} e^{55} e_{\Lambda,5}^{\hphantom{\Lambda,}l} e_{\mu l}  \Big) 
\crbig
&=& \Big( \frac{1}{2}e^{55}e_{\mu,5}^{a} + \frac{1}{2}e^{a\nu} e^{55} e_{\nu,5}^{b} e_{\mu b}  \Big) 
\crbig
&=& e^{55}\,\Big( \partial_5 e^{-\frac{C}{\sqrt{3}}y} \Big)   \Big(\frac{1}{2} \delta_{\mu}^{a} + e^{\frac{1}{\sqrt{3}}Cy}\eta^{\nu \kappa}\delta^a_{\kappa} \delta_{\nu}^{b} \eta_{cb} e_\mu^c \Big) =-\frac{C}{\sqrt{3}} \delta_{\mu}^{a} \,.
\end{array}
\eeq
Similarly, we find that
\beq
\omega_5^{\hphantom{\mu}ab} = \omega_5^{\hphantom{\mu}a5}=0  \,.
\eeq

Since $\partial_\mu \epsilon_{i1} =0$, we have that (in the vacuum) on the brane
\beq
D_\mu \epsilon_{i}=   \frac{1}{4} \omega_\mu^{mn} \Gamma_{mn} \epsilon_{i}  =-\frac{C}{2\sqrt{3}} \Gamma_\mu \Gamma_5 \epsilon_{i}  \,.
\eeq
Then, using the first of the equations (\ref{comp}), the first of the equations (\ref{susytr}) takes the following form on the brane
\beq
\tilde\delta \psi_{\mu i}=\frac{i}{2\sqrt 3} \Gamma_\mu \Big( i C \Gamma^5  \epsilon_{i}  + \frac{gA}{\sqrt2} \varepsilon_{ij}\delta^{jk}\epsilon_k  \Big) \,,
\eeq
while  the $5$--th component of the first of the equations (\ref{susytr}) takes the form
\beq
\tilde\delta \psi_{5 i}= \partial_5 \epsilon_{i} + \frac{igA}{2 \sqrt{6}}\Gamma_5  \varepsilon_{ij}\delta^{jk}\epsilon_k\,.
\eeq


\newpage
\begin{thebibliography}{99}

\bibitem{Antoniadis:2011qw}
  I.~Antoniadis, A.~Arvanitaki, S.~Dimopoulos and A.~Giveon,
  ``Phenomenology of TeV Little String Theory from Holography,''
  Phys.\ Rev.\ Lett.\  {\bf 108} (2012) 081602
  doi:10.1103/PhysRevLett.108.081602
  [arXiv:1102.4043 [hep-ph]].

\bibitem{Baryakhtar:2012wj}
  M.~Baryakhtar,
  ``Graviton Phenomenology of Linear Dilaton Geometries,''
  Phys.\ Rev.\ D {\bf 85} (2012) 125019
  doi:10.1103/PhysRevD.85.125019
  [arXiv:1202.6674 [hep-ph]].
  
\bibitem{Cox:2012ee}
  P.~Cox and T.~Gherghetta,
  ``Radion Dynamics and Phenomenology in the Linear Dilaton Model,''
  JHEP {\bf 1205} (2012) 149
  doi:10.1007/JHEP05(2012)149
  [arXiv:1203.5870 [hep-ph]].
  

  
 \bibitem{Antoniadis:1999rm}
  I.~Antoniadis and B.~Pioline,
  ``Low scale closed strings and their duals,''
  Nucl.\ Phys.\ B {\bf 550} (1999) 41
  doi:10.1016/S0550-3213(99)00151-0
  [hep-th/9902055].

 \bibitem{Antoniadis:2001sw}
  I.~Antoniadis, S.~Dimopoulos and A.~Giveon,
  ``Little string theory at a TeV,''
  JHEP {\bf 0105} (2001) 055
  doi:10.1088/1126-6708/2001/05/055
  [hep-th/0103033].


\bibitem{Giudice:2016yja}
  G.~F.~Giudice and M.~McCullough,
  ``A Clockwork Theory,''
  JHEP {\bf 1702} (2017) 036
  doi:10.1007/JHEP02(2017)036
  [arXiv:1610.07962 [hep-ph]].

\bibitem{Choi:2015fiu}
  K.~Choi and S.~H.~Im,
  ``Realizing the relaxion from multiple axions and its UV completion with high scale supersymmetry,''
  JHEP {\bf 1601} (2016) 149
  doi:10.1007/JHEP01(2016)149
  [arXiv:1511.00132 [hep-ph]].

\bibitem{Kaplan:2015fuy}
  D.~E.~Kaplan and R.~Rattazzi,
  ``Large field excursions and approximate discrete symmetries from a clockwork axion,''
  Phys.\ Rev.\ D {\bf 93} (2016) no.8,  085007
  doi:10.1103/PhysRevD.93.085007
  [arXiv:1511.01827 [hep-ph]].


\bibitem{Aharony:2004xn}
  See e.g. O.~Aharony, A.~Giveon and D.~Kutasov,
  ``LSZ in LST,''
  Nucl.\ Phys.\  B {\bf 691}, 3 (2004)
  [arXiv:hep-th/0404016];
  and references therein.

\bibitem{Aharony:1998ub}
  O.~Aharony, M.~Berkooz, D.~Kutasov and N.~Seiberg,
  ``Linear dilatons, NS5-branes and holography,''
  JHEP {\bf 9810}, 004 (1998)
  [arXiv:hep-th/9808149].


\bibitem{Randall:1999ee}
  L.~Randall and R.~Sundrum,
  ``A Large mass hierarchy from a small extra dimension,''
  Phys.\ Rev.\ Lett.\  {\bf 83} (1999) 3370
  doi:10.1103/PhysRevLett.83.3370
  [hep-ph/9905221].
  %%CITATION = doi:10.1103/PhysRevLett.83.3370;%%
  %7692 citations counted in INSPIRE as of 14 Oct 2017


\bibitem{Gunaydin:1983bi}
  M.~Gunaydin, G.~Sierra and P.~K.~Townsend,
  ``The Geometry of N=2 Maxwell-Einstein Supergravity and Jordan Algebras,''
  Nucl.\ Phys.\ B {\bf 242} (1984) 244.
  doi:10.1016/0550-3213(84)90142-1

\bibitem{Gunaydin:1984pf}
  M.~Gunaydin, G.~Sierra and P.~K.~Townsend,
  ``Vanishing Potentials in Gauged $N=2$ Supergravity: An Application of Jordan Algebras,''
  Phys.\ Lett.\  {\bf 144B} (1984) 41.
  doi:10.1016/0370-2693(84)90172-2

\bibitem{Gunaydin:1984ak}
  M.~Gunaydin, G.~Sierra and P.~K.~Townsend,
  ``Gauging the d = 5 Maxwell-Einstein Supergravity Theories: More on Jordan Algebras,''
  Nucl.\ Phys.\ B {\bf 253} (1985) 573.
  doi:10.1016/0550-3213(85)90547-4
  
  \bibitem{Antoniadis:1995vz}
  I.~Antoniadis, S.~Ferrara and T.~R.~Taylor,
  ``N=2 heterotic superstring and its dual theory in five-dimensions,''
  Nucl.\ Phys.\ B {\bf 460} (1996) 489
  doi:10.1016/0550-3213(95)00659-1
  [hep-th/9511108].

%\bibitem{Awada:1985ep}
%  M.~Awada and P.~K.~Townsend,
%  ``$N=4$ Maxwell-einstein Supergravity in Five-dimensions and Its SU(2) Gauging,''
%  Nucl.\ Phys.\ B {\bf 255} (1985) 617.
%  doi:10.1016/0550-3213(85)90156-7


  %\cite{Benakli:2014daa}
\bibitem{Benakli:2014daa}
  K.~Benakli,
  ``(Pseudo)goldstinos, SUSY fluids, Dirac gravitino and gauginos,''
  EPJ Web Conf.\  {\bf 71} (2014) 00012
  doi:10.1051/epjconf/20147100012
  [arXiv:1402.4286 [hep-ph]].
  %%CITATION = doi:10.1051/epjconf/20147100012;%%

\bibitem{admp}
I.~Antoniadis, A.~Delgado, C.~Markou and S.~Pokorski,
in preparation.

\bibitem{Kehagias:2017grx}
  A.~Kehagias and A.~Riotto,
  ``The Clockwork Supergravity,''
  arXiv:1710.04175 [hep-th].


\end{thebibliography}
\end{document}